\documentclass[conference]{IEEEtran}
\IEEEoverridecommandlockouts
\usepackage{cite}
\usepackage{amsmath,amssymb,amsfonts}
\usepackage{algorithmic}
\usepackage{graphicx}
\usepackage{textcomp}
\usepackage{xcolor}
\usepackage{hyperref}
\def\BibTeX{{\rm B\kern-.05em{\sc i\kern-.025em b}\kern-.08em
    T\kern-.1667em\lower.7ex\hbox{E}\kern-.125emX}}

\begin{document}

\title{Experiences with Integrating Custos Security Services}

\author{\IEEEauthorblockN{Isuru Ranawaka}
\IEEEauthorblockA{\textit{Indiana University}\\
Bloomington, IN, USA \\
isjarana@iu.edu} \\
\IEEEauthorblockN{Juleen Graham}
\IEEEauthorblockA{\textit{Johns Hopkins University}\\
Baltimore, MD, USA \\
juleengrah@gmail.com} \\
\IEEEauthorblockN{Enis Afgan}
\IEEEauthorblockA{\textit{Johns Hopkins University}\\
Baltimore, MD, USA \\
enis.afgan@jhu.edu}
\and
\IEEEauthorblockN{Samitha Liyanage}
\IEEEauthorblockA{\textit{Indiana University}\\
Bloomington, IN, USA \\
shliyana@indiana.edu} \\
\IEEEauthorblockN{Terry Fleury}
\IEEEauthorblockA{\textit{University of Illinois}\\
Champaign, IL, USA \\
tfleury@illinois.edu} \\
\IEEEauthorblockN{Jim Basney}
\IEEEauthorblockA{\textit{University of Illinois}\\
Champaign, IL, USA \\
jbasney@illinois.edu}
\and
\IEEEauthorblockN{Dannon Baker}
\IEEEauthorblockA{\textit{Johns Hopkins University}\\
Baltimore, MD, USA \\
dannon@jhu.edu} \\
\IEEEauthorblockN{Dimuthu Wannipurage}
\IEEEauthorblockA{\textit{Indiana University}\\
Bloomington, IN, USA \\
dwannipu@iu.edu} \\
\IEEEauthorblockN{Suresh Marru}
\IEEEauthorblockA{\textit{Indiana University}\\
Bloomington, IN, USA \\
smarru@iu.edu}
\and
\IEEEauthorblockN{Alexandru Mahmoud}
\IEEEauthorblockA{\textit{Johns Hopkins University}\\
Baltimore, MD, USA \\
almahmoud@jhu.edu} \\
\IEEEauthorblockN{Yu Ma}
\IEEEauthorblockA{\textit{Indiana University}\\
Bloomington, IN, USA \\
yuma@iu.edu} \\
\IEEEauthorblockN{Marlon Pierce}
\IEEEauthorblockA{\textit{Indiana University}\\
Bloomington, IN, USA \\
marpierc@iu.edu}
}

\maketitle
\pagestyle{plain}

\begin{abstract}
Science gateways are user-facing cyberinfrastructure that provide researchers and educators with Web-based access to scientific software, computing, and data resources.  Managing user identities, accounts, and permissions are essential tasks for science gateways, and gateways likewise must manage secure connections between their middleware and remote resources. The Custos project is an effort to build open source software that can be operated as a multi-tenanted service that provides reliable implementations of common science gateway cybersecurity needs, including federated authentication, identity management, group and authorization management, and resource credential management. Custos aims further to provide integrated solutions through these capabilities, delivering end-to-end support for several science gateway usage scenarios. This paper examines four deployment scenarios using Custos and associated extensions beyond previously described work. The first capability illustrated by these scenarios is the need for Custos to provide hierarchical tenant management that allows multiple gateway deployments to be federated together and also to support consolidated, hosted science gateway platform services. The second capability illustrated by these scenarios is the need to support service accounts that can support non-browser applications and agent applications that can act on behalf of users on edge resources. We illustrate how the latter can be built using Web security standards combined with Custos permission management mechanisms. 

\end{abstract}


\section{Introduction}
Science gateways play a vital role in research cyberinfrastructure by bridging the gap between scientists and research computing and data management infrastructure providers \cite{lawrence2015science}. Science gateways provide user-friendly, domain-specific interfaces to researchers, students, and other users to access a wide variety of scientific computational resources while absorbing complexities such as using job schedulers on diverse high performance computing systems, interacting with storage infrastructure, moving and managing data, executing scientific software, and complying with security and usage policies that arise when using shared scientific computing resources. 

Science gateways can be built using open source software such as Galaxy \cite{afgan2018galaxy}, HUBzero \cite{mclennan2010hubzero}, Tapis \cite{cleveland2020tapis}, Open OnDemand \cite{hudak2018open}, or Apache Airavata \cite{marru2011apache}. These systems have common security requirements, including authentication and authorization, user and group management, and resource credential management. At a higher level of organization, gateway deployments may want to form a federation that allows users of one gateway to move to other deployments or that controls access to common resources. Finally, gateways may be deployed on consolidated hosting environments that provide ``Platform as a Service'' capabilities \cite{pierce2020integrating}. In this case, a single gateway is a tenant of the hosting platform. 

Gateways furthermore are not monolithic applications deployed on a single Web server. They may be composed of multiple distributed components or microservices, and they may incorporate ``edge'' applications, or agents, that are deployed directly on target resources. Managing authentication and authorization in these scenarios is an interesting challenge. 

Custos \cite{ranawaka2020custos} is an open-source security framework, currently part of the Apache Airavata project, developed around the end-to-end security requirements of science gateways. Custos’s core capabilities are identity and user management \cite{christie2020managing}, group and sharing management \cite{nakandala2017apache}, and resource credential management  \cite{kanewala2014credential}; these capabilities are based on significant prior work that has been integrated and updated to work as a standalone service. Custos is designed to be operated as a service separate from its client gateways. Custos is accessed through its API by client gateways using either gRPC or REST calls. A Custos Service deployment is designed to serve the security needs of multiple gateways; each gateway is a tenant. Custos leverages best-of-breed security software wherever possible, building on Keycloak \cite{keycloak} to provide authentication and identity management services, and on HashiCorp Vault \cite{vault} to provide resource credential (``secrets'') management. Custos is integrated with CILogon \cite{CILogon2019}, which provides federated authentication to client gateways.  

This paper examines four different Custos integration scenarios: with the Galaxy Project, with the HathiTrust Research Center’s Analytics Gateway and Data Capsule system, with the Apache Airavata-based Science Gateways Platform as a service system, and with the Airavata Managed File Transfer Service. These scenarios illustrate the use of hierarchically arranged tenants and the use of service accounts that support agents deployed outside the managed middleware deployment. 

\section{Galaxy: Towards World-Wide Federation}

\subsection{Galaxy Overview}
The Galaxy Project \url{https://galaxyproject.org/} offers software and services for analyzing biomedical data via a Web browser \cite{10.1093/nar/gkaa434}. The flagship Galaxy open source application provides a consistent Web interface for thousands of domain-specific command-line tools that have been ``wrapped'' for use with Galaxy and makes those tools easily accessible to domain researchers \cite{goecks2010galaxy}. Once a user submits a job, the job is executed on a suitable HPC or cloud resource and, once complete, the outputs are returned to the user via Galaxy \cite{afgan2011galaxy}. Typically, all jobs are submitted using a generic system user account without regard for the identity of the researcher submitting the job or using the service. The integration of Custos with Galaxy will enable a much richer experience for researchers and more accountable system insights for the service provider by connecting an individual user’s identity to their Galaxy account \cite{afgan2018federated}.

Galaxy can be installed locally or used via any of the many managed Galaxy services provided by the Galaxy community. The most popular Galaxy services are known as usegalaxy.* (e.g., usegalaxy.org, usegalaxy.org.au, usegalaxy.eu). This is a federation of world-wide Galaxy installations hosted on national infrastructure from around the world that adheres to a set of common guidelines. Because each service is hosted on a given nation’s public infrastructure, researchers from that country may be given higher storage quotas or priority in running jobs. For example, usegalaxy.org.au service offers a six-times-larger storage quota for Australian researchers. Currently, this is realized through an ``allow list'' concept that is manually curated based on the domain of the user’s email address, which may be error prone and may change over time. 

\subsection{Galaxy-Custos Integration}
Thus far, Galaxy has developed integration with Custos for handling user authentication. Users are no longer required to create a local username and password but can use any of 4,000+ Identity Providers (IdPs) available via Custos and brokered via CILogon to authenticate. While improving accessibility of the Galaxy service for users, this facilitates service providers to ``reason'' about the user’s identity without manual interpretation of the user’s registration information (e.g., domain of the registered email address). Starting with Galaxy release 21.01, Galaxy supports native integration with Custos where enabling this form of user authentication is a matter of simple configuration. Once enabled, users can login using their institutional credentials while Galaxy, via Custos, can associate the user’s identity with an institution from InCommon and other supported federations. A given Galaxy installation will internally process the now-available user’s affiliation as an authoritative source of information and act accordingly (e.g., automatically provide larger quota or submit jobs to more capable machines). 

The second form of Galaxy integration with Custos is support for resource credentials (secrets) management. Custos offers capabilities to store and manage users' sensitive information in the form of key-value pairs, such as authorization tokens, SSH keys, and passwords for third-party applications. The benefits of storing secrets with Custos include the following: (1) the gateway does not hold this high-risk data; (2) the Custos implementation is validated to ensure proper handling of such data; and (3) Custos stores the data in an encrypted data store based on HashiCorp’s Vault. While Custos provides secure management of these secrets, client applications such as Galaxy can retrieve the secret on behalf of the user by using the user’s consistent identity. Traditionally, Galaxy has refrained from wrapping tools that require users to authenticate using their password because the only mechanism for Galaxy to interact with those remote services was to store the password in plain text in its local database. This password could also leak into the application logs and hence interaction with such services was not enabled in Galaxy. With Custos integration, such secrets can be retrieved on behalf of the user from the Custos resource credentials service without being stored locally. The user would link a secret stored in Custos with their account in Galaxy via user preferences (e.g., CloudStor app passwords or NCBI API keys) and the secret would be used to populate an input field in the tool form (e.g., use a stored password) or to facilitate an authorization flow between Galaxy and the third-party service (e.g., use a stored token), depending on how the given tool is wrapped. 

\begin{figure}[htbp]
  \centering
  \includegraphics[width=\linewidth]{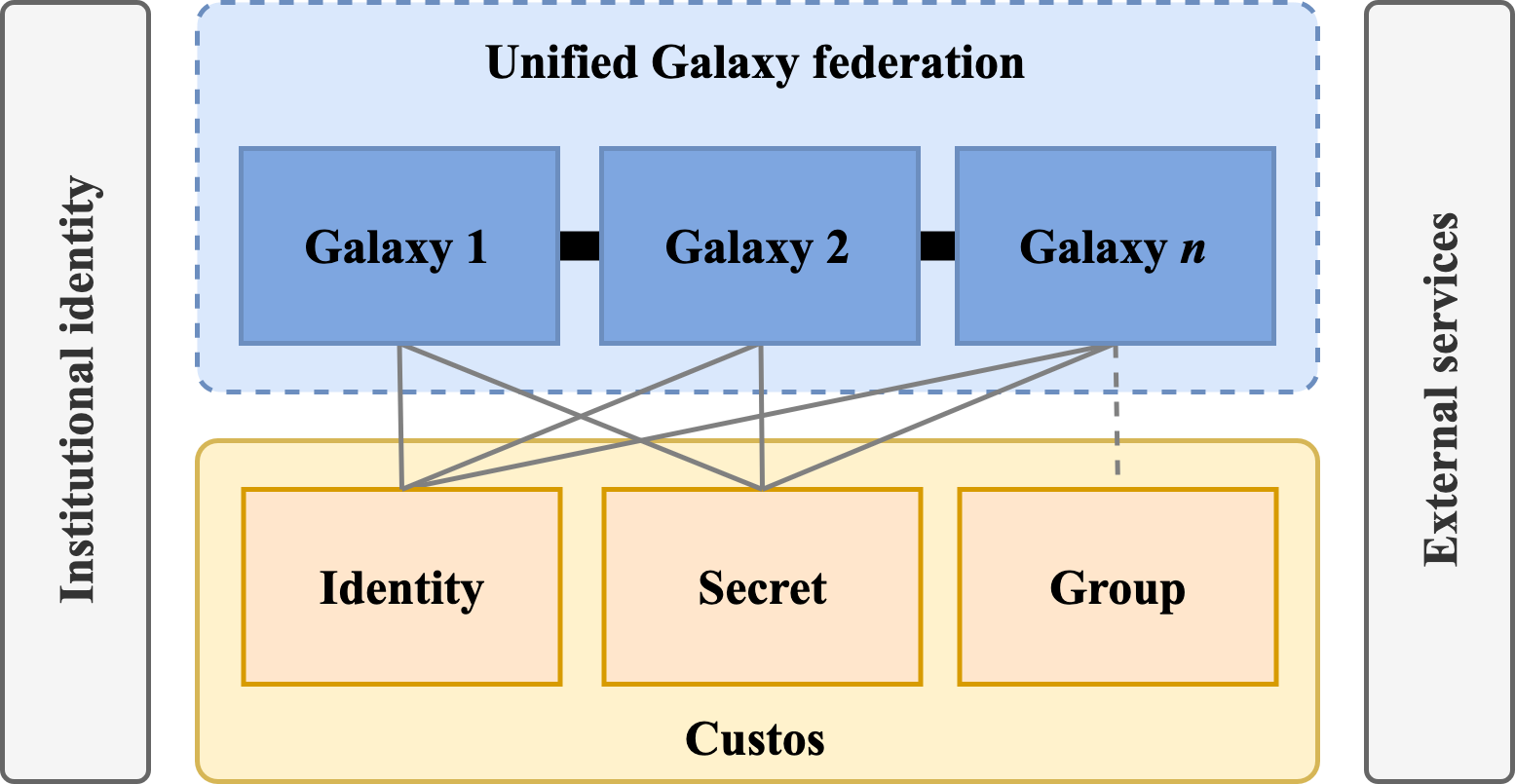}
  \caption{Custos enabling better connectivity and eventual federation of Galaxy services. All Galaxy instances can by default enable Custos identity integration but selectively (based on the need) can choose to take advantage of secret and group management capabilities.}
  \label{figGalaxyIntegrations}  
\end{figure}

Looking ahead, the availability of the user’s stable identity and a centralized location for secrets paves a path for unifying the usegalaxy.* services. Rather than each Galaxy service being a completely independent installation with its own database of users and the user’s data being tied to the single service, Galaxy is on track to provide a single, global Galaxy service that is federated across multiple Galaxy installations and uniformly perceived by users. To start, users will be able to log in using their preferred credentials and retrieve the same secrets from Custos regardless of which usegalaxy.* server they are using. Once support for federated data is enabled in Galaxy, it will be possible to extend this notion to access a user’s data uniformly across any Galaxy instance. Visualized in Fig. \ref{figGalaxyIntegrations}, Custos integration is a critical component of this vision for seamless Galaxy federation. Currently, Custos is facilitating the use of user identities within Galaxy and the use of protected external services. In the future, the interconnectivity of the identity, external services, and Galaxy services will enable the development of a well-connected, federated Galaxy service that links multiple components to deliver an enhanced user experience and provide better tools for service providers. 

\section{HathiTrust Research Center: Diverse Authentication Requirements}
The HathiTrust Research Center (HTRC) enables computational analysis of the HathiTrust Digital Library, which is the largest non-profit digital library in the world. HTRC is a collaborative research center launched jointly by Indiana University and the University of Illinois, along with HathiTrust, which is based at the University of Michigan. HTRC’s mission is to help meet the technical challenges researchers face when dealing with massive amounts of digital text available through HathiTrust. Using the data storage and computational infrastructure at Indiana University and the University of Illinois at Urbana-Champaign, HTRC develops and deploys tools for large-scale text mining and non-consumptive research \cite{htrc_ncrup}, allowing scholars to utilize HathiTrust content fully while preventing intellectual property misuse within the confines of U.S. copyright laws.

HTRC supports users primarily through two mechanisms: a Web-based science gateway front end for basic analysis and an innovative ``data capsule'' system that gives controlled command-line access to advanced users \cite{kowalczyk2016big}. We examine the support for both of these mechanisms with Custos. 

\subsection{Authenticating HTRC Analytics Gateway Users}

The HTRC Analytics Gateway provides Web-based access to tools for analyzing the HathiTrust collection while preventing users from directly accessing restricted materials. The organizational separation between HathiTrust and HTRC introduces some authentication and authorization challenges. Some HathiTrust services are limited to individuals affiliated with those in its membership community, which consists primarily of academic institutions. HathiTrust uses federated identities to access these digital library services, while the HathiTrust Research Center has maintained its own unique identity provider. Due to this, users of services from both organizations must authenticate separately to each service, and cross-organizational services are impossible due to disconnected identities. Members of HathiTrust often view the digital library and HTRC services as part and parcel of the larger HathiTrust enterprise, and find the separate authentication systems confusing.

\begin{figure}[htbp]
  \centering
  \includegraphics[width=\linewidth]{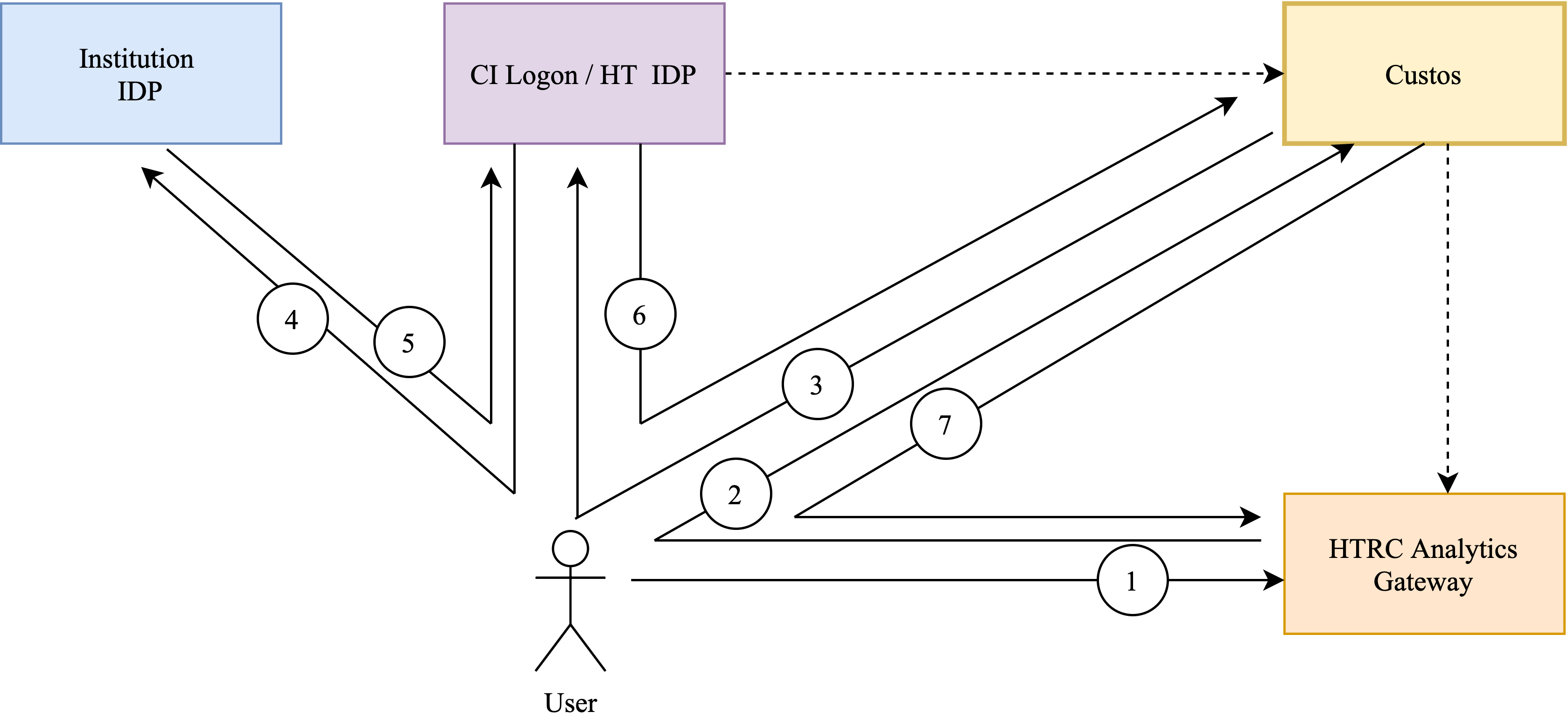}
  \caption{Unifying identity management between HathiTrust and the HTRC Analytics Gateway.}
  \label{figHTRCUnifying}  
\end{figure}

A revised Identity and Access Management (IAM) infrastructure for HTRC aims to solve the first problem by brokering identities via Custos for authenticating to HTRC managed services. Information flow is depicted in Fig. \ref{figHTRCUnifying}. A user (1) goes to the HTRC Analytics Gateway and selects his/her institution from the dropdown list to initiate authentication. The HTRC Analytics Gateway (2) sends the request to the Custos server with the correct external IdP Hint \cite{aarc_idp_hint} and the selected institution’s entityID as parameters. Custos (3) sends the request to the external IdP (either CILogon or HathiTrust’s own IdP) with the institution’s entityID as a parameter. The external IdP (4) redirects the user to the login page of the user’s institution. The user (5) provides institutional credentials and authenticates with the institution's identity server. Successful authentication responses return to the external IdP with an access token and an id\_token. The external IdP (6) forwards the authentication response and access tokens to the Custos server. Finally, Custos (7) returns the authentication response to HTRC Analytics Gateway with access tokens. Failed authentications return error codes along the same path.

In this scenario, Custos supports two external identity providers, CILogon and HathiTrust. CILogon federates all educational institutions supporting InCommon and eduGAIN, but some HathiTrust users come from universities outside these federations. HathiTrust operates an instance of OpenAthens \cite{openAthens} as an Open ID Connect (OIDC) endpoint for these users. Additionally, support for the HathiTrust identity provider results in a more consistent login experience for HathiTrust members across the HathiTrust and HTRC services, as described above. Custos registers aforementioned identity providers under different aliases and HTRC requests a particular alias to select a specific identity provider from Custos. HTRC keeps institutions and identity provider mapping to avoid duplicate account creation from two different identity providers.

\subsection{HTRC Data Capsule Service Accounts}

HTRC Data Capsules complement the HTRC Analysis Gateway by providing individual, secure computing environments that allow users to undertake more open-ended analysis of the content in the HathiTrust Digital Library. Researchers can create virtual machines (a.k.a., ``capsules'') where they can import and then analyze HathiTrust text data. Researchers can only perform computational analysis within the secure Data Capsule environment and then export the results of their analysis. Volume text may not be exported outside the HTRC Data Capsule, and data products leaving a capsule must undergo review prior to release to ensure they meet the HTRC's policy for non-consumptive data exports. All users may access public domain items. Computational access to items in copyright is available only to HathiTrust member-affiliated researchers due to resource constraints.

Data Capsules are created through the HTRC Analytics Gateway as shown in Fig. 3. Users are authenticated through the process described in the preceding section. After the capsule is created, a user loads the desired volumes onto the virtual machine using the HTRC Data API using command line tools within the capsule. API calls are secured using the capsule-specific ID tokens that are retrieved from the token service running in the host. 

HTRC uses Custos service accounts (described in greater detail below) for this purpose. Each Data Capsule is represented by a Custos service account, and service account credentials can be used with OAuth2’s Client Credentials grant type  \cite{hardt2012oauth}.

\begin{figure}[htbp]
  \centering
  \includegraphics[width=\linewidth]{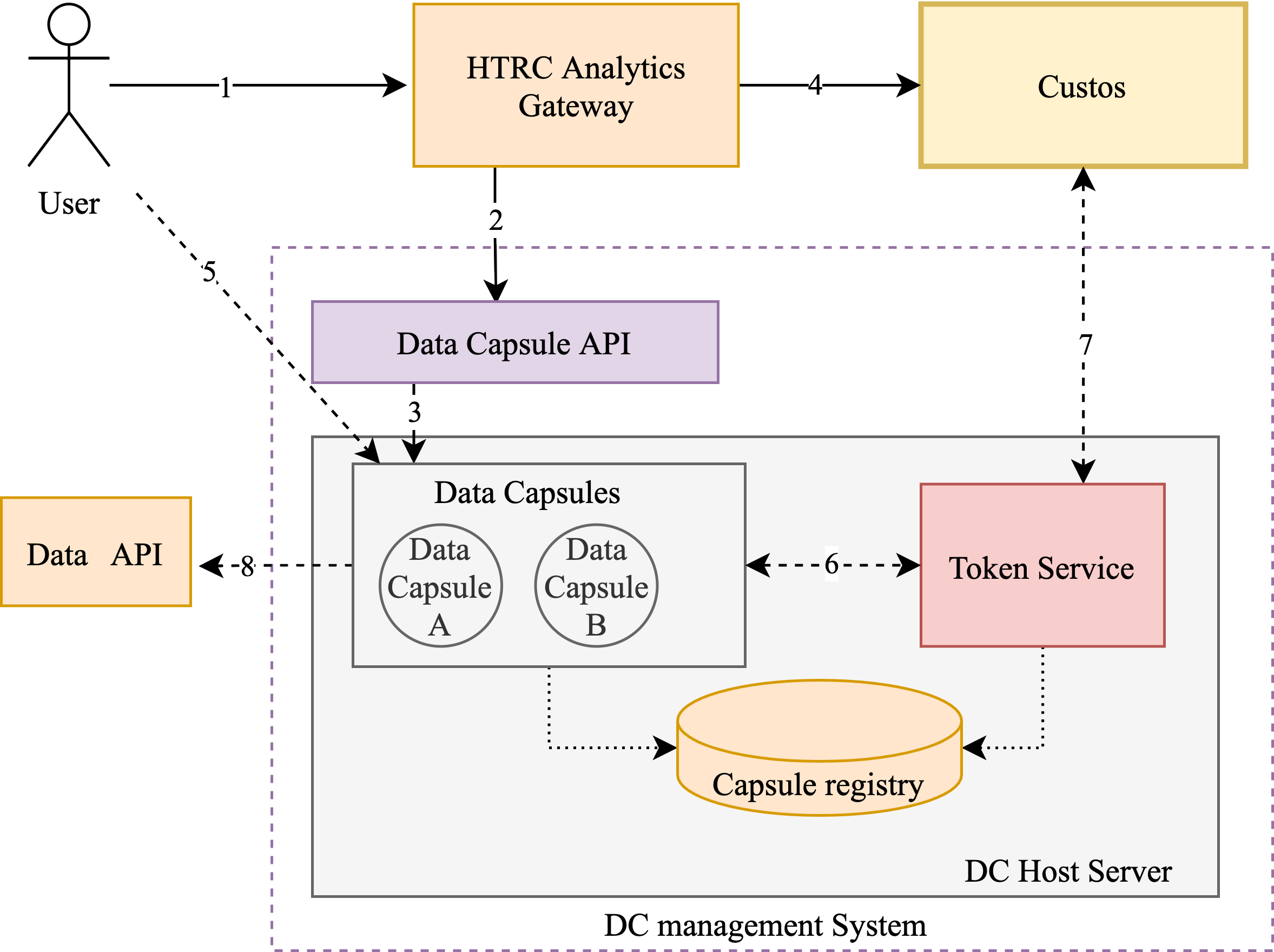}
  \caption{Capsule agent creation and token retrieval steps.}
  \label{figHTRCCapsule}  
\end{figure}

Fig. \ref{figHTRCCapsule} shows the capsule-specific agent creation process. An authenticated user (1) requests the Analytics Gateway to create a Data Capsule. The HTRC Analytics Gateway (2) sends a request to the Data Capsule API and creates a capsule. The Data Capsule service creates a unique capsule ID that is included in the Data Capsule API response. The Data Capsule service (3) deploys the created Data Capsule on the DC Host server. The Analytics Gateway (4) uses the capsule ID and sends a request to Custos to create a service account for the capsule. Custos generates a service account ID and secret that it returns to the gateway. Custos will delete this service account when the capsule is deleted. The Analytics Gateway sends a request to the Data Capsule API to add service account credentials into the capsule’s configuration file stored in the capsule registry.

An authenticated user (5) connects to the Data Capsule in the secure mode and requests to download volumes using HTRC WorksetToolkit, a Python tool. The Data Capsule (6) sends a request to the Token Service deployed in the Data Capsule host. This service validates requests against the capsule ID and capsule’s internal IP. The Token Service (7) validates the request, gets service account credentials from the capsule configuration file and sends a token request to Custos using a client credential grant type. The Token Service (6) sends the id\_token from the token response it received from Custos to the Data Capsule. HTRC WorksetToolkit (8) uses the id\_token received from the Token Service and sends a Data API request.

\subsection{HTRC Service Accounts}

As described previously, we have chosen OAuth2’s Client Credentials grant flow as the mechanism for authenticating Data API requests from the WorksetToolkit command line tools executed by a user within a Data Capsule virtual machine. This token flow, unlike the more common Authorization grant type, does not require mechanisms built into the HTTP protocol that are suitable for browser-based users. OAuth2’s Resource Owner Password Credentials grant type is another option, but this involves the use of long-term credentials rather than access tokens. 

\begin{figure}[htbp]
  \centering
  \includegraphics[width=\linewidth]{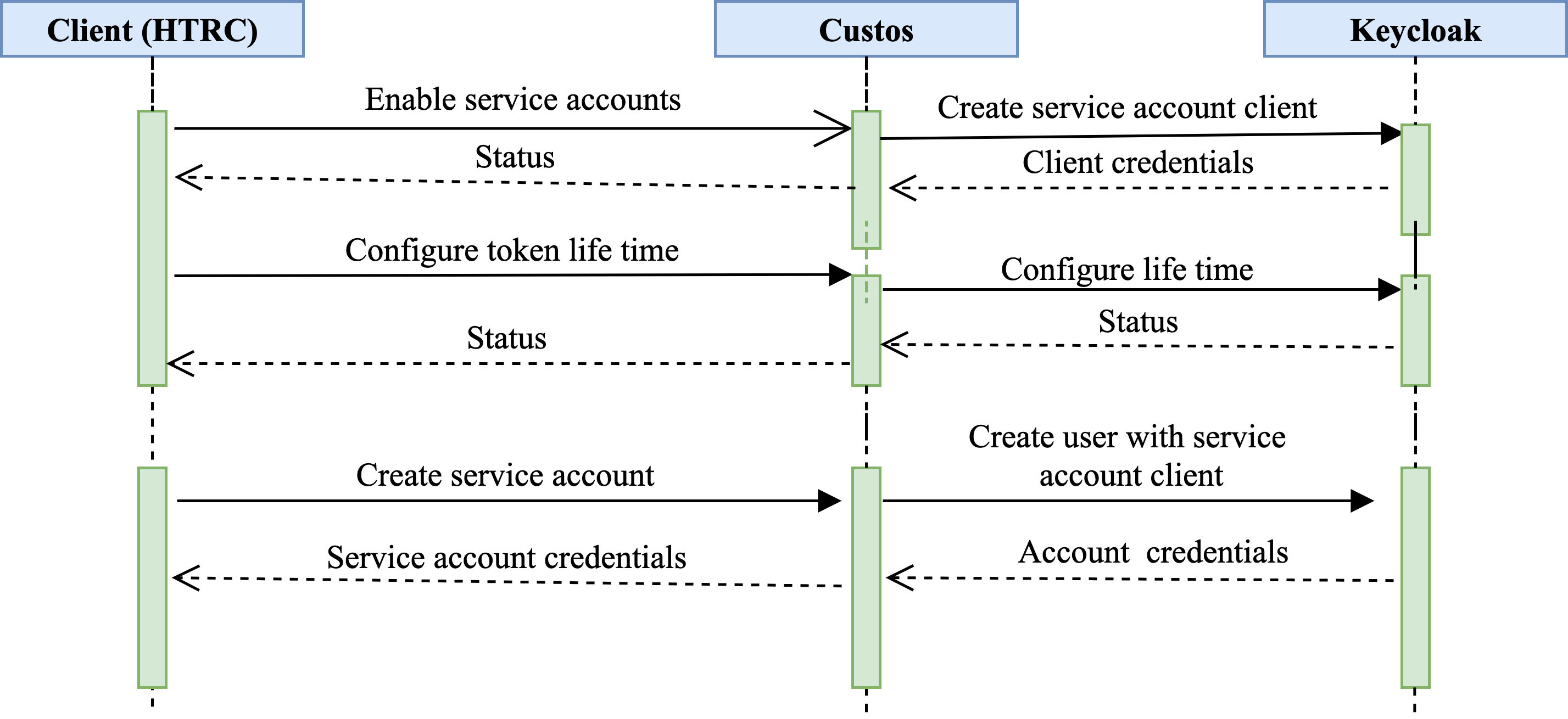}
  \caption{Service Account Registration}
  \label{figServiceAccounts}  
\end{figure}

Supporting Client Credentials grant flows introduces the requirement in Custos to support service accounts. Service accounts are suitable for non-browser processes that need to authenticate to resources on the user’s behalf. Custos provides service accounts to register and authenticate the aforementioned Data Capsules. HTRC’s Analytics Gateway can register service accounts under its tenant using its primary credentials. The main service needs to provide the service account name, roles, and attributes at registration. Once registered, Custos creates service account credentials that can be used to authenticate with Custos in subsequent calls. Fig. \ref{figServiceAccounts} depicts the service account registration message flow.

Authentication flow and the token lifetime of service accounts are managed through separate OAuth clients distinct from user management OAuth clients to provide fine-grained authentication management.

\section{Apache Airavata Platform Integration: Hierarchical Tenant Management}
The Cyberinfrastructure Integration Research Center at Indiana University operates the Science Gateways Platform as a service (SciGaP.org) \cite{pierce2018supporting}, a managed deployment of the Apache Airavata software, to provide a hosted solution for over forty science gateway tenants. A gateway provider can request a gateway tenant through the SciGaP.org administrative portal. Once approved, a SciGaP administrator provisions gateways requested through the portal and provides domain URLs for the newly deployed gateway. These operations can be categorized into two major steps: gateway creation and gateway configuration. This extended example illustrates how Custos’s tenant concepts can be integrated into a platform service where individual science gateways or other clients act as tenants.

\begin{figure}[htbp]
  \centering
  \includegraphics[width=\linewidth]{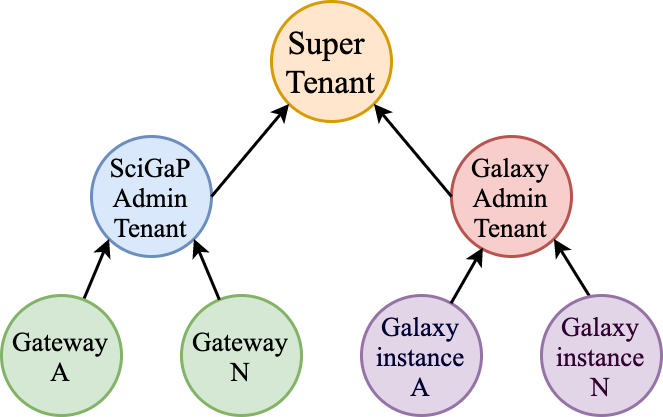}
  \caption{Custos tenants can be hierarchies.}
  \label{figTenantHierarchies}  
\end{figure}

The key concept introduced here is that platform services require a hierarchy of tenants within a centrally operated Custos Service. A platform service such as SciGaP or a Galaxy federation acts as a top level tenant that, in turn, manages its own tenants within Custos. This is depicted in Fig \ref{figTenantHierarchies}. 

\begin{figure}[htbp]
  \centering
  \includegraphics[width=\linewidth]{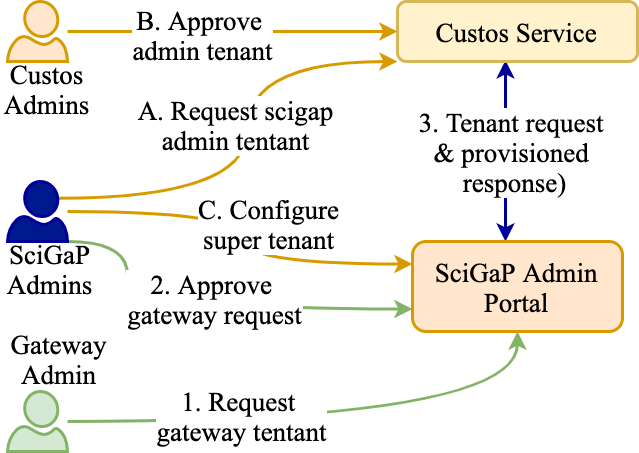}
  \caption{SciGaP super tenant provisioning use case.}
  \label{figSciGaPProv}  
\end{figure}

Fig. \ref{figSciGaPProv} shows the sequence of operations involved in the gateway tenant management process for SciGaP. As a single step, SciGaP administration portal (A) is registered with Custos as an administrator tenant. Administrator tenants (B) are manually approved by the Custos administrator. Once approved, the SciGaP administrator (C) can download the SciGaP primary credentials from Custos and configure SciGaP to use these in subsequent tenant creation requests for new gateways. Subsequent gateway tenant requests are not required to be approved by Custos and onus is on SciGaP admins to validate and approve gateway tenant. When a gateway requester (1) requests a new gateway tenant and SciGaP administrator reviews and approves the request (2), the SciGaP administration portal communicates with Custos (3) using previously-registered credentials to create Custos tenants. Custos considers this request an authorized request and automatically configures Keycloak and CILogon for the new tenant. All operations are processed automatically. Once completed, Custos will send the new Custos credentials for the newly requested tenant to the SciGaP administration portal. The SciGaP administrator can use these credentials to configure a Django Portal for Airavata \cite{christie2020extensible} instance for the newly created gateway; gateway requesters can also implement their own clients using the Apache Airavata API. Once the gateway tenant becomes available, its users can use CILogon-based federated authentication and other identity solutions enabled by Custos.

Once a tenant gateway is created and registered with Custos, it can use other Custos services through Custos’s client API. In addition to authentication via CILogon, the new gateway tenant can use Custos to provide identity management, authentication and authorization services, user and group management, service accounts, and resource credential management services to provide identity management solutions for gateways.

\section{Using Custos Service Accounts to Manage Agent Applications}

The Airavata Managed File Transfer (MFT) service \cite{mftGateways20} is an open-source managed file transfer framework. It is designed to be operated both as an extension to Apache Airavata middleware in a deployment such as SciGaP, and as a standalone service that exposes its own API to client gateways and other applications. Airavata MFT can be extended to work with multiple transfer protocols and different types of storage endpoints, including user-provided cloud storage end-points as well as traditional HPC and mass storage devices. 

\begin{figure}[htbp]
  \centering
  \includegraphics[width=\linewidth]{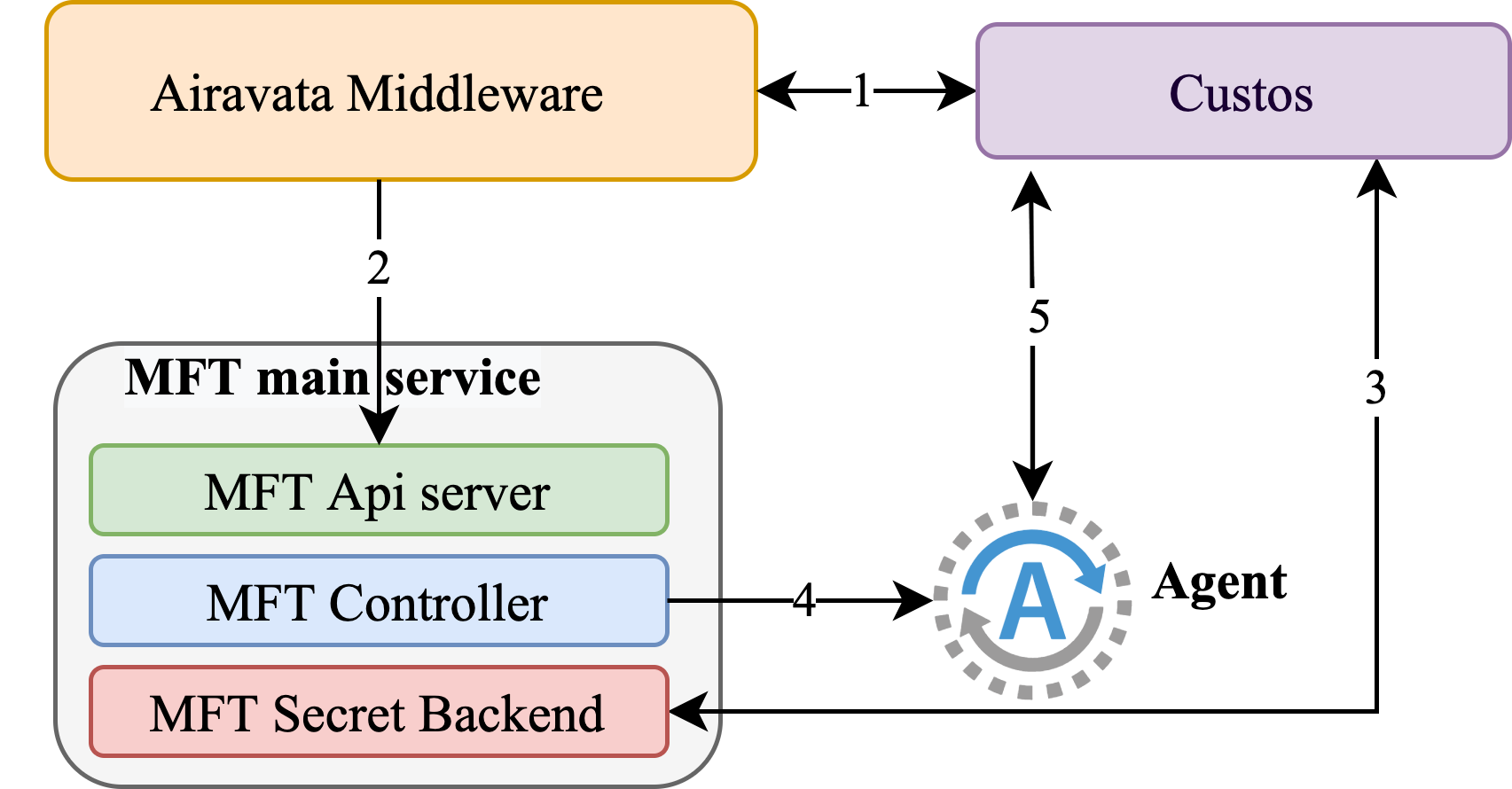}
  \caption{Airavata, MFT, and Agent interaction with Custos.}
  \label{figMFTCustos}  
\end{figure}

Airavata MFT deployments can include agent programs that run directly on storage endpoints as well as centrally-operated middleware; these agents can be used to route data intelligently and efficiently without directly going through central MFT service deployments. Apache Airavata middleware utilizes the MFT framework to copy data from the original data source to the destination data location for a given job execution. Accessing data repositories on remote clusters requires authentication and authorization. Hence, MFT needs access to the login credentials of heterogeneous remote accounts. Moreover, these accounts have diverse types of resource credentials such as SSH keys and OAuth2 access tokens. The MFT framework has a pluggable security framework that can use the Custos resource credential service to store those accounts’ credentials. Custos’s resource credentials management service issues credential tokens for each stored credential to access them later; these are stored in the client application. The MFT framework’s distributed deployment model is depicted in Fig. \ref{figMFTCustos}. 

Managing authorization for agents is also an essential element of the MFT service: agents are used to move data between accounts that may have different owners, such as a user’s Google Drive account and a community account on an XSEDE resource. Each request has different privilege levels and scopes. In terms of the OAuth2 specification \cite{hardt2012oauth}, the MFT central service and MFT client gateways or middleware are Web applications: they are deployed in secured environments and so can have full privileges to fetch credentials from Custos. Agents, on the other hand, correspond to the specification’s definition of native applications. They can be deployed in external environments that are not directly under the control of the service operator; hence, agents should have restricted privileges to access Custos. 

Moreover, agents should have access credentials of particular users' storage and remote clusters to transfer data in between. Agents can obtain these credentials directly through Custos by authenticating themselves with Custos, or the MFT central service can obtain these credentials from Custos and forward them to requested agents. We have introduced OAuth-based access control schemes known as agent-based access control, user-based access control, and delegated access control based on credential retrieval patterns to accommodate different scenarios.

\subsection{Agent-based access control}

\begin{figure}[htbp]
  \centering
  \includegraphics[width=\linewidth]{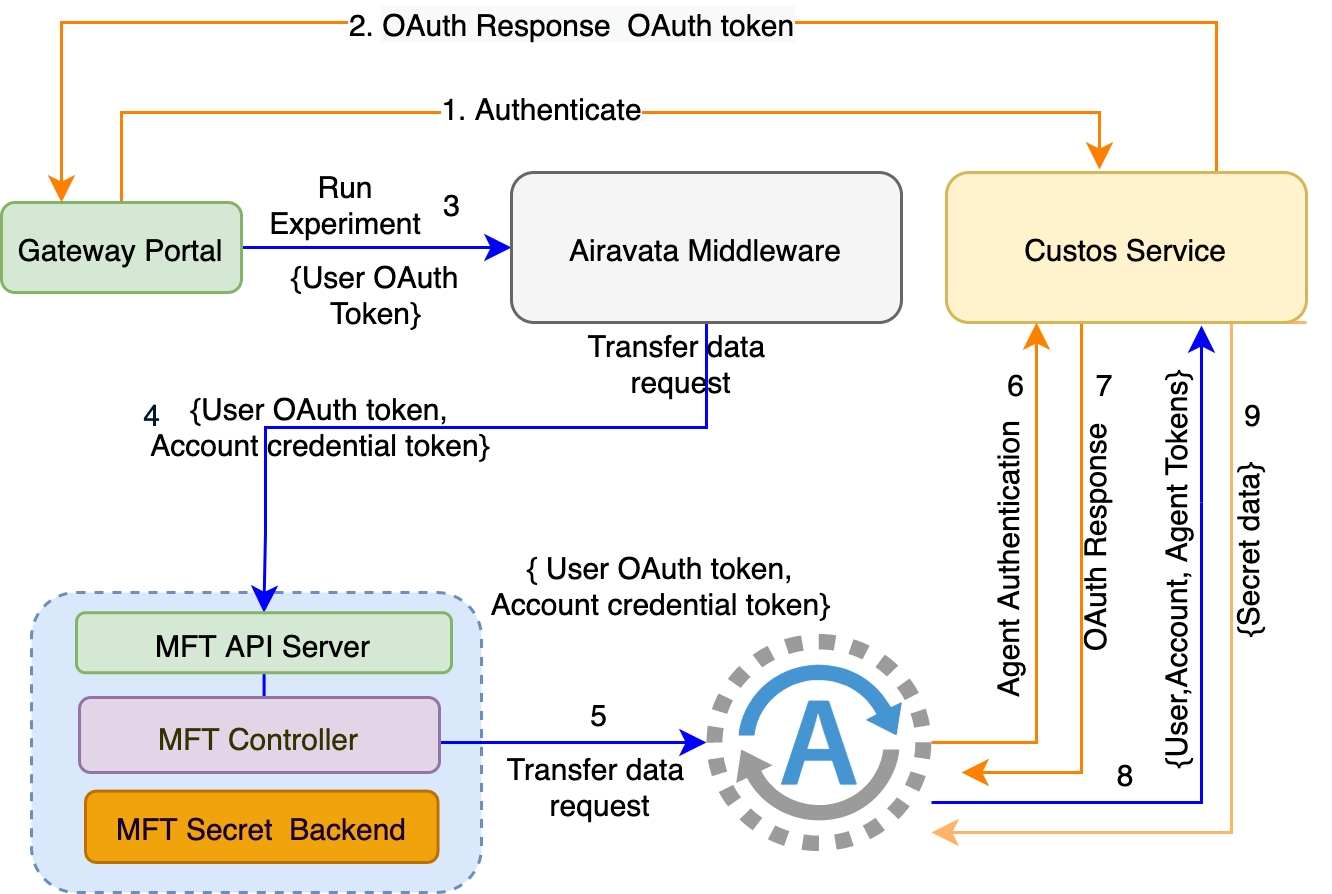}
  \caption{Agent-based access control.}
  \label{figAgentBasedAC}  
\end{figure}

In this scenario, the agent application is directly registered in Custos as a tenant and can request credentials directly to perform tasks for users. As shown in Fig. \ref{figAgentBasedAC}, the user authenticates through the gateway portal, retrieves OAuth tokens from Custos, and initiates a data transfer request. Airavata middleware sends credential tokens required to fetch particular credentials from Custos to initiate the data transfer. The central MFT service forwards the credential token to agents to fetch credentials from Custos and initiate the data transfer. To this end, an agent first authenticates with Custos using an OAuth2 client credential grant type and obtains access tokens. Second, it fetches required credentials from Custos by sending an agent token and a credential token. Custos validates the agent token and identifies the tenant to which it belongs. Furthermore,  Custos identifies requested credentials through the credential token and validates permissions to access requested credentials through the agent token. If permissions are granted, requested credentials are delivered to the agent and the agent will initiate the data transfer.

\subsection{Delegating access control}
\begin{figure}[htbp]
  \centering
  \includegraphics[width=\linewidth]{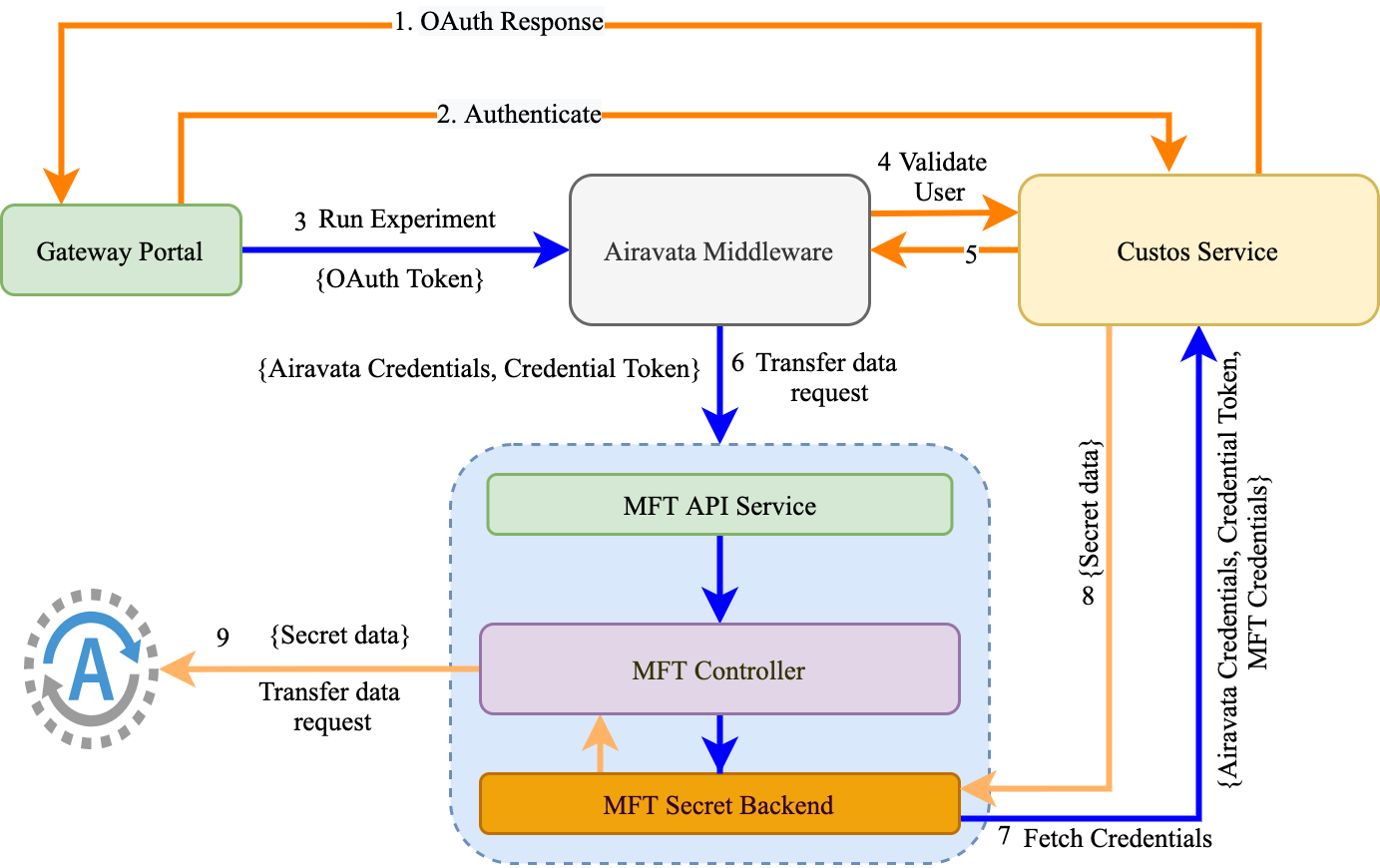}
  \caption{Delegating access control.}
  \label{figDelegatingAC}  
\end{figure}

 The delegating access control scheme is designed to fetch credentials directly from Custos for a previously authorized service or entity; see Fig. \ref{figDelegatingAC}. For instance, the Airavata middleware is a Custos-registered service that can authenticate and check the authorization status of user tokens from Custos before forwarding requests to the MFT service. In this scenario, MFT does not need to reauthorize the user tokens; it fetches relevant credentials directly from Custos only by using Airavata credentials and credential tokens. Subsequently, it delivers credentials to the selected agent to initiate the data transfer. The requested credential needs to be shared with relevant permissions to access through Airavata credentials.
 
\subsection{User-based access control}
\begin{figure}[htbp]
  \centering
  \includegraphics[width=\linewidth]{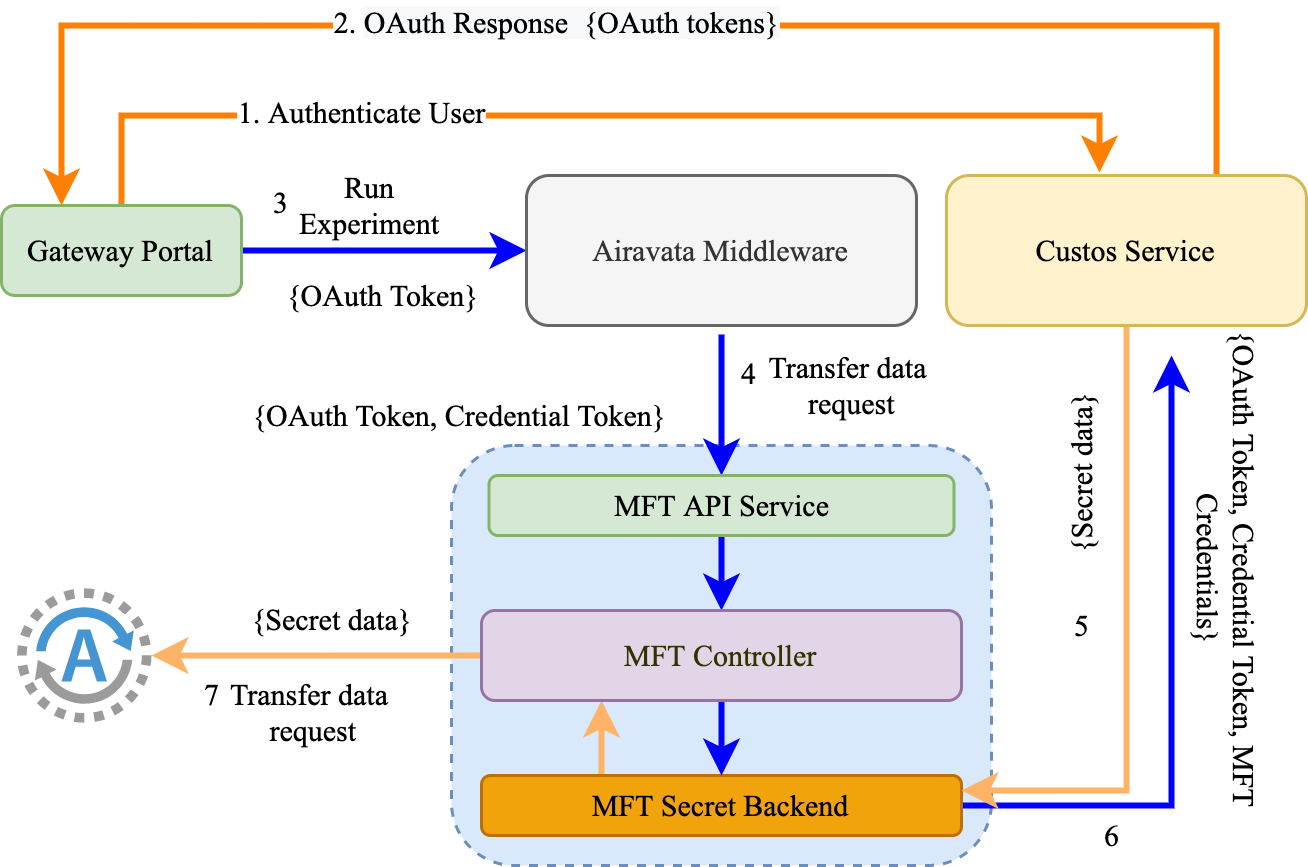}
  \caption{User-based access control.}
  \label{figUserBasedAC}  
\end{figure}

The user-based access control scheme is designed to fetch credentials directly from Custos using a user’s OAuth tokens; see Fig. \ref{figUserBasedAC}. MFT uses the user’s tokens directly to fetch  relevant credentials from Custos and subsequently  delivers them to the selected agent to initiate the data transfer. In this scenario, credentials do not need to be shared with any external service such as Airavata, MFT or agents to be fetched from Custos. User credentials and tokens are passed opaquely through Airavata middleware. 

Agent-based access control can be used for scenarios in which agents are deployed in trustworthy environments. Delegating access control will be useful to initiate multiple data transfers through trusted middleware services without reauthenticating users. User-based access control is recommended for scenarios in which credentials cannot be shared with external entities or services.

While this section has focused on Apache Airavata-MFT agent integration scenarios, it is possible to generalize these to other types of middleware or to other gateway providers in future work. 

\section{Custos Administrative Portal}

\begin{figure}[htbp]
  \centering
  \includegraphics[width=\linewidth]{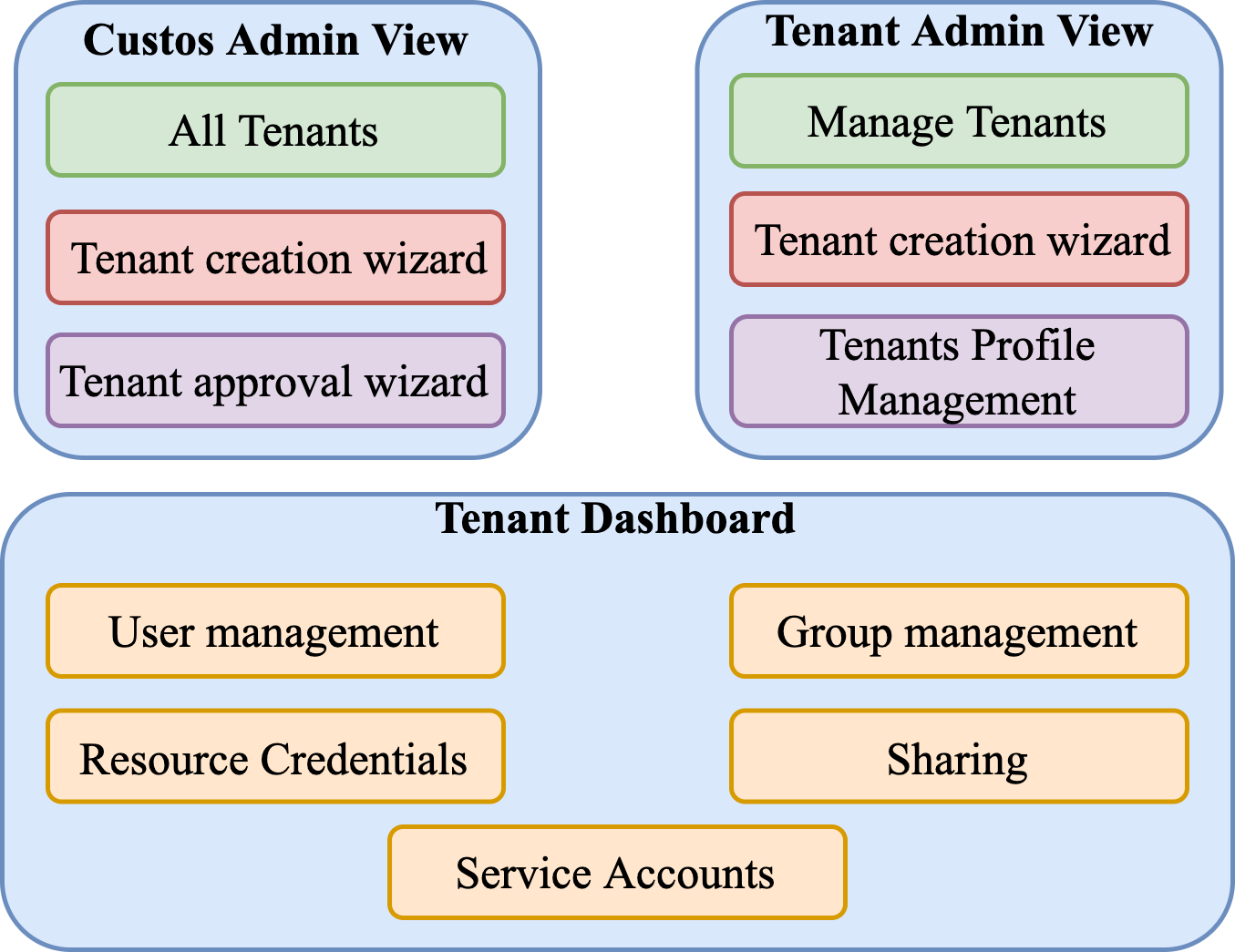}
  \caption{Overview of Custos Admin Portal Capabilities.}
  \label{figAdminPortalOverview}  
\end{figure}

The Custos administration Portal supplements REST, gRPC interfaces and convenient Custos SDKs. The portal provides capabilities to request and manage tenants. Key capabilities as illustrated in Fig. \ref{figAdminPortalOverview} include Tenant requester dashboard, Custos admin dashboard, and Tenant management dashboard. 

\begin{figure}[htbp]
  \centering
  \includegraphics[width=\linewidth]{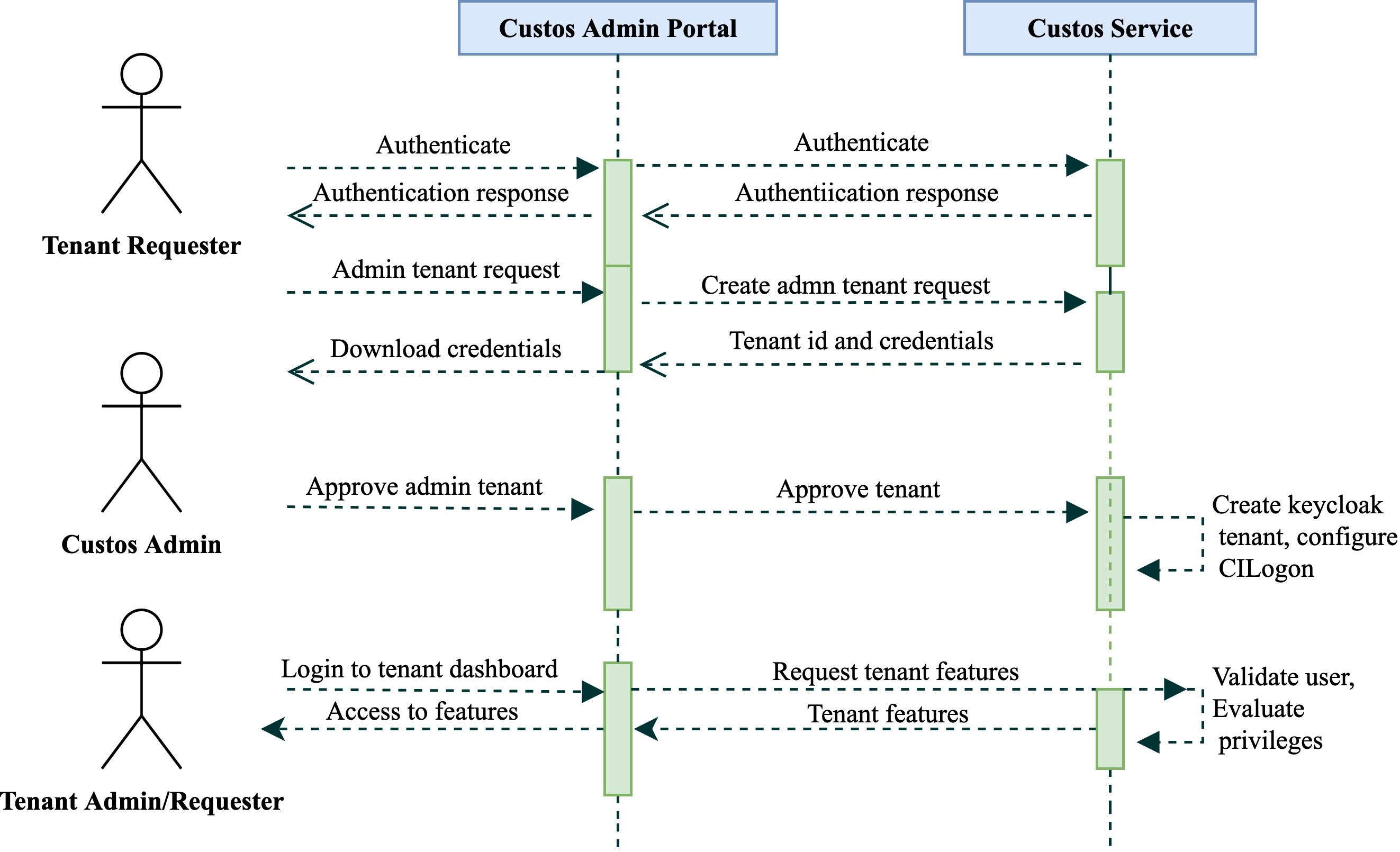}
  \caption{Sequence of Admin Portal Operations.}
  \label{figAdminPortalSequence}  
\end{figure}

Fig. \ref{figAdminPortalSequence} illustrates the sequence of operations from tenant requesting to tenant approval and tenant login. A tenant requester follows a guided wizard to request a Custos tenant. Custos administrators use the portal to view all the tenant requests and profiles, approve or reject tenant requests, and create tenants. Upon approval and tenant creation, tenant administrators can manage tenant configurations. Tenants can utilize the option of enabling users, and can also use the portal for managing user profiles, groups and secrets. For user-level operations, we anticipate gateway tenants consume API's and build in all user management functionality into their respective gateway interfaces, but the Custos admin portal provides extra convenience, if it is a needed capability.

\section{Related Work}
Tapis \cite{cleveland2020tapis} provides API’s for building Science Gateways. Like Custos, Tapis Security Kernel supports OpenID Connect and OAuth for authentication and uses Vault for storing secrets. As discussed in this paper, Custos is a general purpose security infrastructure while Tapis security kernel is integrated within the Tapis infrastructure for internal specific usage.  D4Science \cite{D4Science} is a virtual research environment hosting platform that supports federated authentication via the European Open Science Cloud (EOSC) and authorization attribute-based access control. D4Science uses the gCube software framework \cite{gCube}, similar to SciGaP's use of Airavata and Custos. A key difference is that Custos emphasizes the management of credentials for connecting to external HPC and cloud resources, whereas D4Science uses computing resources internal to the D4Science infrastructure. Galaxy's CloudAuthz \cite{galaxyCloudAuthz} provides federated identity and access management for accessing biomedical datasets across multiple cloud platforms, using OpenID Connect and OAuth. The CloudAuth approach requires the setup of a service role on the cloud provider so that it grants access for the application (i.e., Galaxy) on behalf of the user. Once configured, the use of this flow is limited to the services and tools that support it. The Galaxy integration with Custos Secrets broadens interaction with protected services by allowing their use via passwords or tokens without requiring support for access roles by the service provider. In \cite{ranawaka2020custos} we reviewed Custos comparison with Globus, COmanage and Grouper. 

Keycloak provides multi-tenant federated identity and access management services, including support for OpenID Connect, OAuth, and SAML. Keycloak's fine-grained authorization services support attribute-based and role-based access control. Custos builds on Keycloak, adding interfaces that are customized to science gateway use cases, along with secrets management using Vault. 

\section{Conclusion and Future Work}
Custos is open source software for operating a service that provides an integrated approach to authentication, user management, group and role management, and security credential management that can be used to implement multiple end-to-end scenarios for science gateways and other cyberinfrastructure. When operated as a service, such as the Custos Service operated by the Cyberinfrastructure Integration Research Center, Custos can manage a hierarchy of security realms, called tenants. In the scenarios described in this paper, a tenant is a particular gateway deployment that is either integrated into a federation of other gateways that share services (such as a Galaxy federation) or integrated into a centralized platform deployment such as SciGaP. This is an extension over our original tenant concept \cite{ranawaka2020custos}. 

The other major requirement illustrated in this paper is using Custos to manage service accounts, such as for HTRC’s Data Capsule system and for Airavata MFT agents. Science gateways are typically Web browser-based and so can use Web protocols such as OAuth2’s Authorization Code Grant flow, which in turn is based on the operation of the Web server in a secure environment. Service accounts, in contrast, are suitable for non-Web browser clients (such as command-line tools and agent applications). OAuth2’s Client Credential Grant flow is designed for these scenarios. Custos implements both of these grant flows.

\section*{Acknowledgment}

This work is supported by NSF Award \#1840003.

\bibliographystyle{IEEEtran}
\bibliography{bib-custos-escience21}

\end{document}